\documentclass[conference, 10pt, a4paper]{IEEEtran}
\usepackage{cite}
\usepackage{amsmath,amssymb,amsfonts, amsthm}
\usepackage{algorithmic}
\usepackage{graphicx}
\usepackage{paralist}
\newtheorem{proposition}{Proposition}
\usepackage{url}

\newcommand{\KeyGen}{\pcalgostyle{KeyGen}}
\newcommand{\AESCBC}{\pcalgostyle{AES\text{-}CBC}}

\usepackage[n,advantage,operators,sets,adversary,probability,notions,primitives,events,complexity,keys]{cryptocode}

\newtheorem{definition}{Definition}
\usepackage{algorithm}
\begin{document}

\title{Empirical Ciphertext Distinguishability Testing of KEMs and Hybrid Encryption via Deep Learning

}

\author{
  \IEEEauthorblockN{Simon Calderon\textsuperscript{1, 2} and Niklas Johansson\textsuperscript{1, 2} and Onur G\"unl\"u\textsuperscript{3, 1}}
    \IEEEauthorblockA{\textsuperscript{1}%
        Department of Electrical Engineering, Link{\"o}ping University, Sweden
    }
    \IEEEauthorblockA{\textsuperscript{2}%
        Sectra Communications, Link{\"o}ping, Sweden
    }
    \IEEEauthorblockA{\textsuperscript{3}%
        Lehrstuhl f{\"u}r Nachrichtentechnik, Technische Universit{\"a}t Dortmund, Germany\\
        simon.calderon@liu.se, niklas.johansson@liu.se, onur.guenlue@tu-dortmund.de
    }
}

\maketitle

\begin{abstract}
We model indistinguishability under chosen-plaintext attack (IND-CPA) games as binary classification tasks for deep neural networks (DNNs). We apply this methodology to the public-key encryption (PKE) schemes used to construct PQC key encapsulation mechanisms (KEMs) such as ML-KEM and HQC, and to cascaded symmetric algorithms, demonstrating methodological generality. Moreover, we present a novel extension of this approach to empirical distinguishability testing of hybrid KEMs. For IND-CPA-secure schemes and their combinations, no DNN achieves accuracy significantly different from random guessing (measured via Bonferroni-corrected binomial tests), consistent with theoretical security guarantees. For insecure schemes, such as plain RSA, AES-CTR and AES-CBC with IV or counter reuse, or RC4, the DNNs classify the ciphertexts better than random guessing. Thus, this framework is shown to serve as a practical, scheme-agnostic tool for flagging distinguishable ciphertexts in both individual and composed constructions.
\end{abstract}


\section{Introduction}
Development and validation of cryptographic algorithms require significant effort. Careful consideration of the security assumptions and rigorous cryptanalysis are needed before an algorithm is employed to secure information. Recently, the development of new algorithms has been driven by the threat of quantum computers running Shor's algorithm \cite{shor1994algorithms}, which could be used to efficiently attack widely deployed public-key primitives, such as RSA \cite{rivest1978method} and Diffie-Hellman \cite{diffie1976new}. This led the National Institute of Standards and Technology (NIST) \cite{NIST-PQC} to initiate a standardization process for post-quantum cryptography (PQC) algorithms that resulted in the standardization of ML-KEM \cite{FIPS203}, a key encapsulation mechanism (KEM) for key exchange, among other algorithms. Due to long replacement times for cryptographic primitives and ``harvest now, decrypt later attacks'', the transition to PQC is currently of high priority \cite{mosca2022quantum}. 

One technique used to facilitate the adoption of PQC is called hybrid encryption \cite{bindel2019combiners}. Hybrid methods combine several existing algorithms so that the combined algorithm is secure as long as at least one of the component algorithms is secure. This allows PQC and classical algorithms to be combined to benefit from the cryptanalysis performed and security assumptions established for each algorithm, thus facilitating faster adoption by spreading the risk.

The connections between cryptography and machine learning methods were mentioned as early as 1991 by Rivest \cite{rivest1991cryptography}, drawing a parallel between them as two cases in which one tries to learn an unknown function. Neural networks have since been used for key recovery against DES and 3DES \cite{Alani3DES} and to construct distinguishers for reduced-round Speck \cite{gohr2019} and other lightweight ciphers such as Ascon \cite{baksi2022machine}. Recently, Kim et al. \cite{kim2025cryptanalysismachinelearningbased} demonstrated the use of Deep neural networks (DNNs) for cryptanalysis, which includes modeling IND-CPA games as binary classification tasks for a DNN. We extend their method to establish deep learning-based IND-CPA results on both PQC KEMs and hybrid methods that combine PQC and classical cryptographic methods to provide an adaptive testing method for complex hybridization methods. 

Specifically, we extend a DNN-based approach used for IND-CPA testing to demonstrate its applicability to testing PQC KEMs and further extend it to test hybrid KEMs. The proposed neural-network framework complements formal security analysis by providing a practical and flexible empirical tool to evaluate indistinguishability and identify potential weaknesses in implementations and composed constructions. The main contributions of this work include:
\begin{itemize}
    \item We use DNN-based modeling of the IND-CPA game to empirically test the IND-CPA properties of the underlying PKE schemes used in ML-KEM and HQC.
    \item We present a novel extension to this framework to test hybridized KEMs, allowing for testing hybridized KEMs, where the KEM is constructed using the XOR combiner $k=k_1\oplus k_2$ \cite{Giacon2018Combiners}, where $\oplus$ is bitwise XOR, $k_1 $ and $k_2$ are shared secrets from the component KEMs, and $k$ is the resulting shared secret, by combining any KEM with any Public Key Encryption (PKE).
    \item We test and demonstrate the methodological generality of our deep learning-based constructions by applying them also to other hybridized scenarios, such as cascade symmetric encryption \cite{schneier2007applied}.
    \item For the above experiments, we introduce rigorous statistical analysis through two-sided binomial hypothesis testing to determine whether the DNN classifiers gain a significant advantage. We also use chi-squared testing as a baseline classifier to demonstrate the advantage of using DNN classifiers. 
\end{itemize}
\section{Preliminaries}\label{sec:Prel}

\emph{Notation}:
$x \sample  S$ denotes the uniform sampling of $x$ from a set $S$,  and define $[a:b]:=\{n\in \mathbb Z \mid a\leq n\leq  b\}$. We use $\theta$ to denote the parameters of a DNN. For DNN training, we denote the actual class of the sample $i$ by $Y_i$ and the network's predicted output by $\hat{Y}_i$, respectively.

\subsection{Basic Definitions}

We now define the main security notion used. While the following definition is formulated for PKEs, the corresponding definition for symmetric schemes is analogous (under appropriate modifications) and is omitted for brevity.

\begin{definition}[IND-CPA {\cite[p.~401]{katz2025introduction}}]
    For an encryption scheme $\Pi$, an adversary $\mathcal A$ and a security parameter $\lambda$, the experiment $\mathrm{PubK}^{\mathrm{cpa}}_{\mathcal A, \Pi}(\lambda)$ is defined by the following game: A challenger $\mathcal C$ uses $\Pi.\mathrm{keygen}(1^\lambda)$ to generate a key pair $(pk,sk)$ and gives $pk$ to the adversary $\mathcal A$. The adversary then selects two messages $(m_0,m_1)$ of equal length and gives them to the challenger. The challenger proceeds by selecting a bit $b$ uniformly at random and handing the adversary $\Pi.\mathrm{enc}(pk, m_b)$. The adversary is tasked with outputting a bit $b'$. The output of the experiment is $1$ if $b=b'$ and 0 otherwise.
     An encryption scheme $\Pi$ has indistinguishable ciphertexts under chosen plaintext attack if for all probabilistic polynomial-time adversaries $\mathcal A$ there exists a negligible function $\varepsilon(\lambda)$ such that
    \begin{equation}
        \Pr\Big[\mathrm{PubK}^{\mathrm{cpa}}_{\mathcal A, \Pi}(\lambda)= 1\Big] \leq \frac{1}{2}+\varepsilon(\lambda).
    \end{equation}\hfill $\lozenge$
\end{definition}

\begin{definition}[Binary cross entropy (BCE) \cite{goodfellow2016deep}]
BCE loss can be used to quantify the discrepancy between the true labels and the approximated distributions produced by the network. Specifically, the average BCE over a batch of size $N$ is used. Let $Y_i\in\{0,1\}$ be the actual label of sample $i$, and $\hat{Y}_i\in (0,1)$ be the probability $\Pr[Y_i=1]$ predicted by the network. Then, the batch loss is given by 
\begin{equation}\label{eq:BCE_batch}
    \mathrm{BCE}(Y,\hat{Y})=-\frac{1}{N}\sum_{i=1}^N [Y_i\log(\hat{Y}_i)+(1-Y_i)\log(1-\hat{Y}_i)].
\end{equation}\hfill $\lozenge$
\end{definition}

\subsection{Hybrid Encryption}\label{subsec:HybridEncryption}
Combining cryptographic algorithms dates back to Shannon \cite{shannon1949communication}. Several potential goals motivate such combinations: creating stronger encryption than single algorithms,  hedging against potential weaknesses in any individual algorithm, and providing fail-safes against misconfigurations.

\subsubsection{Hybridized KEMs}
We first consider the hybridization of KEMs.

\begin{definition}[KEM Combiners {\cite[p.~198]{Giacon2018Combiners}}]\label{def:kem_comb}
      Let $k_1,\dots ,k_n$ be the output from $n$ KEMs, and let $c_1,\dots, c_n$ be their corresponding ciphertexts. A KEM combiner of $n$ KEMs is a function $W$ such that $k= W(k_1,\dots, k_n, c_1, \dots, c_n)$, where $k$ is the output shared secret. \hfill $\lozenge$
    
\end{definition}

For the rest of this work, we focus exclusively on the XOR combiner, as defined next.

\begin{definition}[XOR combiner{\cite[p.~198]{Giacon2018Combiners}}]\label{def:ps_comb}
     Let $k_1,k_2\in \mathcal K=\{0,1\}^\ell$ be the output shared secrets from two KEMs. We have $k=k_1\oplus k_2$, where $k$ is the output shared secret. With independently generated component keys and independent encapsulation randomness, this combiner is IND-CPA secure if at least one component KEM is IND-CPA secure. \hfill $\lozenge$
\end{definition}

\subsubsection{Hybridization of Symmetric Encryption}
For two symmetric encryption algorithms $\mathrm{Enc}_\mathrm a$ and $\mathrm{Enc}_\mathrm b$, cascade encryption is defined as $\mathrm{Enc}_{\mathrm{cascade}}(m)=\mathrm{Enc}_\mathrm a(\mathrm{Enc}_\mathrm b(m))$ \cite{schneier2007applied}.

Let $\mathrm{Enc}_a$ be the outer cipher and $\mathrm{Enc}_b$ the inner cipher.
\section{Deep Learning, KEM, and Cascade Encryption Setups for IND-CPA Testing}\label{sec:DLandKEM}
In this section, we present the DNN-based IND-CPA testing methodology, our algorithm selections, network configurations, and statistical evaluation methods.

\subsection{The IND-CPA Testing Algorithm for PKEs and Cascade Encryption}
We employ Algorithm \ref{alg:bce_cpa} from \cite{kim2025cryptanalysismachinelearningbased}, which is given below, to model IND-CPA games using DNNs. For each algorithm, a DNN is trained to distinguish between two classes of ciphertext, one class consisting of encryptions of uniformly random plaintexts, and another class consisting of encryptions of a fixed plaintext, in this case, all 0s. Each DNN is trained using stochastic gradient descent to minimize Binary Cross-Entropy (BCE) loss. If the encryption scheme used is IND-CPA secure, any polynomial-time training-and-classification procedure has at most a negligible advantage over random guessing, averaged over key generation and all procedure randomness.

We remark that Algorithm \ref{alg:bce_cpa} is not directly applicable to all KEMs, for example, KEMs constructed with Fujisaki--Okamoto transform or variants thereof, since Algorithm \ref{alg:bce_cpa} requires selecting a plaintext, which is not possible for such constructions. However, Algorithm \ref{alg:bce_cpa} can be applied to test the IND-CPA security of the PKE schemes that are used in the Fujisaki--Okamoto style transform to create the KEM. The motivation to test this is that the IND-CPA security of the underlying PKE is used to prove the IND-CCA2 security of the KEM \cite{fujisaki1999secure}. Moreover, Algorithm \ref{alg:bce_cpa} is used for IND-CPA testing of cascade symmetric algorithms.

\begin{algorithm}[htpb]
\caption{IND-CPA BCE Classification adversary \cite{kim2025cryptanalysismachinelearningbased}}\label{alg:bce_cpa}
\begin{algorithmic}
\STATE \textbf{Input} Plaintext set $X_0 \sample [0 : 2^{\ell}-1]^N$ to challenger.
\STATE \textbf{Input} Plaintext set $X_1 \gets \{0^\ell\}^N$ to challenger.
\STATE \textbf{Challenger} outputs ciphertexts sets $Y_{0}$ and $Y_{1}$ from $X_{0}$ and $X_{1}$
\STATE \textbf{Form labeled data} $(y,b)$ with $y\in Y_b$, $b \in \{0,1\}$
\STATE Initialize network parameters $\theta$
\REPEAT 
    \STATE Evaluate $\mathbf{BCE}(Y,\hat{Y})$ on the true labels and the predicted scores
    \STATE Compute SGD optimizing and updating $\theta$
\UNTIL{convergence}
\end{algorithmic}
\end{algorithm}

\subsection{Relation to Other Security Notions}
 We note that the DNN adversary differs from the standard IND-CPA adversary in two key aspects: it receives only ciphertexts as input, without access to the public key or an encryption oracle, and its strategy is learned from a fixed training dataset rather than through adaptive queries. Consequently, the DNN adversary is strictly weaker than a standard IND-CPA adversary. If a DNN achieves accuracy significantly above chance, this constitutes evidence of a distinguishable pattern that a standard IND-CPA adversary could also exploit. Conversely, failure to distinguish does not constitute a proof of IND-CPA security, but means that the tested procedures did not detect a statistically significant classification advantage.

One-wayness under chosen plaintext attack (OW-CPA) is in the conventional hierarchy the next step down from IND-CPA. The following examples show that while our DNN adversary is strictly weaker than the standard IND-CPA adversary, OW-CPA and resistance to the restricted ciphertext-only classification task considered in this work do not imply each other.

\begin{definition}[OW-CPA \cite{HofHeinz2017AModular}] 
Let $\Pi = (\mathsf{KeyGen}, \mathsf{Enc}, \mathsf{Dec})$ be a public-key encryption scheme with message space $\mathcal{M}$, and let $\mathcal{A}$ be an adversary and $\lambda$ a security parameter. The experiment $\mathsf{PubK}^{\mathsf{ow\text{-}cpa}}_{\mathcal{A},\Pi}(\lambda)$ is defined as follows:
\begin{inparaenum}
    \item The challenger runs $(pk, sk) \leftarrow \mathsf{KeyGen}(1^\lambda)$ and gives $pk$ to $\mathcal{A}$.
    \item The challenger samples $m \sample \mathcal{M}$ uniformly at random, computes $c \leftarrow \mathsf{Enc}_{pk}(m)$, and gives $c$ to $\mathcal{A}$.
    \item $\mathcal{A}$ outputs $m'$.
    \item The output of the experiment is $1$ if $m' = m$, and $0$ otherwise.
\end{inparaenum}
The scheme $\Pi$ is \emph{one-way under chosen-plaintext attack} if for all probabilistic polynomial-time adversaries $\mathcal{A}$ there exists a negligible function $\varepsilon(\lambda)$ such that
\begin{equation}
    \Pr\left[\mathsf{PubK}^{\mathsf{ow\text{-}cpa}}_{\mathcal{A},\Pi}(\lambda) = 1\right] \leq \varepsilon(\lambda).
\end{equation} \hfill $\lozenge$
\end{definition}

\subsubsection*{OW-CPA does not guarantee resistance against the DNN adversary}
Consider the following scheme. Let $\Pi=(\mathsf{KeyGen},\mathsf{Enc},\mathsf{Dec})$ be some OW-CPA PKE scheme. Now define the following scheme $\tilde{\Pi}$ as specified in Figure \ref{fig:leaky_ow_cpa}.

\begin{figure}[b]
\centering
\small
\begin{pchstack}[boxed, space=1.5em]
  \procedure{$\tilde\Pi.\mathsf{Enc}(pk,m)$}{%
    c \gets \Pi.\mathsf{Enc}(pk,m) \\
    \pcif m = 0^\ell \pcthen b \gets 0 \ \pcelse b \gets 1 \\
    \pcreturn (b,c)}
  \procedure{$\tilde\Pi.\mathsf{Dec}(sk,(b,c))$}{%
    \pcreturn \Pi.\mathsf{Dec}(sk,c)}
\end{pchstack}
\caption{The scheme $\tilde\Pi$, which reveals whether $m = 0^\ell$ or not. $\tilde{\Pi}.\mathsf{KeyGen}$ is the same as $\Pi.\mathsf{KeyGen}$}
\label{fig:leaky_ow_cpa}
\vspace{-0.3cm}
\end{figure}

For uniformly sampled messages in $\{0,1\}^{\ell(\lambda)}$, a classifier using the first bit achieves accuracy $1-2^{-\ell-1}$. Simulating the extra bit as $1$ changes any inverter's success probability by at most $2^{-\ell}$. Thus $\tilde{\Pi}$ remains OW-CPA provided $2^{-\ell(\lambda)}$ is negligible.

\subsubsection*{Resistance against the DNN adversary does not guarantee OW-CPA} Consider the PKE $\Pi_{AES}$, defined in Figure \ref{fig:asymmetric_aes}. This scheme is not OW-CPA, as an adversary holding the public key can use it as the secret key to decrypt any challenge ciphertext, recovering the entire plaintext. To the DNN adversary, this is not classifiable, as it interacts with this scheme as it would with the ordinary $\mathsf{AES\text{-}CBC}$ scheme, which is an IND-CPA scheme. Hence, the DNN IND-CPA adversary cannot be compared to the OW-CPA security notion.

\begin{figure}[b]
\vspace{-0.5cm}
\centering
\begin{pchstack}[boxed, space=1.5em]
  \procedure{$\Pi_{AES}. \KeyGen()$}{%
    \pk \sample \mathcal{K} \\
    \sk \gets \pk \\
    \pcreturn (\pk, \sk)
  }
  \pchspace
  \begin{pcvstack}
    \procedure{$\Pi_{AES}.  \enc_{\pk}(m)$}{%
      c \gets \AESCBC.\enc(\pk, m) \\
      \pcreturn c
    }
    \procedure{$\Pi_{AES}. \dec_{\sk}(c)$}{%
      m \gets \AESCBC.\dec(\sk, c) \\
      \pcreturn m
    }
  \end{pcvstack}
\end{pchstack}
\caption{An asymmetric version of AES-CBC that is not OW-CPA, whose ciphertext-only view matches symmetric AES-CBC testing with a hidden key.}
\label{fig:asymmetric_aes}
\vspace{-0.5cm}
\end{figure}

\subsection{The IND-CPA Testing Algorithm for hybrid KEMs using the XOR combiner}
The methodology of Algorithm~\ref{alg:bce_cpa} does not translate to hybrid KEMs, since the hybrid KEM  does not allow for the selection of plaintext messages. For a scenario in which the XOR combiner is used to combine a KEM with a PKE, the methodology can be adapted as presented in Algorithm~\ref{alg:xor_combiner_ind_cpa}. In these experiments, the DNN receives as input the concatenation of the ciphertexts from the two components. The corresponding binary classification task distinguishes between (i) samples where the two ciphertext components decrypt/decapsulate to the same underlying value (the KEM shared secret), and (ii) samples where they decrypt/decapsulate to two values of the same length, generated independently of each other.

\begin{algorithm}[htpb]
\caption{IND-CPA BCE Classification adversary for hybrid KEM}\label{alg:xor_combiner_ind_cpa}
\begin{algorithmic}
\STATE \textbf{Input} Shared secret set $X_{0} $, and ciphertext set $Y_{A0}$ from $N$ encapsulations with the KEM.
\STATE \textbf{Input} Ciphertext set $Y_{A1}$ from $N$ encapsulations with the KEM. 
\STATE \textbf{Input} plaintext set  $X_1 \sample [0:2^{\ell}-1]^N$, where $\ell$ is the KEM shared secret length.
\STATE \textbf{Challenger} outputs ciphertext sets $Y_{B0}$ and $Y_{B1}$ from $X_{0}$ and $X_{1}$ using the PKE.
\STATE \textbf{Challenger} forms joint ciphertexts $Y_{0} = Y_{A0}\|Y_{B0}$ and $Y_{1} = Y_{A1}\|Y_{B1}$ by pairwise concatenating elements
\STATE \textbf{Form labeled data} $(y,b)$ with $y\in Y_b$, $b \in \{0,1\}$
\STATE Initialize network parameters $\theta$.
\REPEAT 
    \STATE  
    Evaluate $\mathbf{BCE}(Y,\hat{Y})$ on the true labels and the predicted scores
    \STATE Compute SGD optimizing and updating $\theta$
\UNTIL{convergence}

\end{algorithmic}
\end{algorithm}

The rationale behind Algorithm~\ref{alg:xor_combiner_ind_cpa} is the following XOR identity: If $k_1=k_2$, then $k_1 \oplus k_2= 0^{\ell}$. When $k_1$ and $k_2$ are sampled independently and uniformly from $\mathcal{K}$, equality occurs with probability $\Pr[k_1=k_2]=1/|\mathcal{K}|$. This creates two structurally different cases, which extend and are analogous to the IND-CPA modeling viewpoint in Algorithm~\ref{alg:bce_cpa}, where a classifier is trained to separate two label classes corresponding to different underlying conditions. The classification task in Algorithm~\ref{alg:xor_combiner_ind_cpa} tests whether the combined ciphertexts $c_1\|c_2$ leak any information that can be utilized by our neural network architectures to distinguish between the combined key being $0^\ell$ and a key being derived from independent components. This distinction concerns the induced PKE defined below; it does not by itself establish an attack on the standard shared-secret indistinguishability game of an independently encapsulated XOR-combined KEM.

\subsection{Interpretation of Algorithm \ref{alg:xor_combiner_ind_cpa}}
In this subsection, we demonstrate that the hybrid setup is the standard IND-CPA setup for an Induced PKE, defined in the following manner. 

\begin{definition}[Induced PKE]\label{def:induced}
Let $\mathsf{KEM}$ be a KEM with shared-secret space $\{0,1\}^\ell$ and let
$\mathsf{PKE}$ be a PKE whose message space contains $\{0,1\}^\ell$. The
\emph{induced PKE} $\Pi'=\Pi'[\mathsf{KEM},\mathsf{PKE}]$, with message space
$\{0,1\}^\ell$, one-time-pads the message with the KEM's shared secret and
encrypts the result under $\mathsf{PKE}$ (Fig.~\ref{fig:induced}). \hfill $\lozenge$
\end{definition}

\begin{figure}[t]
\centering
\small
\begin{pchstack}[boxed]
  \procedure{$\Pi'.\mathsf{Enc}(pk,m)$}{%
    (c_1,s) \gets \mathsf{KEM}.\mathsf{Encaps}(pk_1) \\
    c_2 \gets \mathsf{PKE}.\mathsf{Enc}(pk_2,\, m\oplus s) \\
    \pcreturn (c_1,c_2)}
  \pchspace
  \procedure{$\Pi'.\mathsf{Dec}(sk,(c_1,c_2))$}{%
    s \gets \mathsf{KEM}.\mathsf{Decaps}(sk_1,c_1) \\
    \pcreturn \mathsf{PKE}.\mathsf{Dec}(sk_2,c_2)\oplus s}
\end{pchstack}
\caption{The induced PKE $\Pi'$. $\Pi'.\mathsf{KeyGen}$ runs both key
generations and outputs $pk=(pk_1,pk_2)$ and $sk=(sk_1,sk_2)$.}
\label{fig:induced}
\vspace{-0.3cm}
\end{figure}

\begin{proposition}\label{prop:alg2}
Up to exchanging class labels, Algorithm~\ref{alg:xor_combiner_ind_cpa} has the ciphertext distributions of Algorithm~\ref{alg:bce_cpa} applied to $\Pi'$: class~0 consists of encryptions of $0^\ell$, and class~1 of encryptions of uniformly random $m\in\{0,1\}^\ell$ under $\Pi'$

\end{proposition}

\begin{proof}
For $m=0^\ell$, $\Pi'.\mathsf{Enc}$ outputs $(c_1,\mathsf{PKE}.\mathsf{Enc}(pk_2,s))$,
which is a class-0 sample. For $m$ uniform and independent of $(c_1,s)$, the
value $m\oplus s$ is uniform and independent of $(c_1,s)$, so the output is
distributed exactly as a class-1 sample.
\end{proof}

Consequently, the interpretation of Algorithm~\ref{alg:bce_cpa} carries over:
a polynomial-time training-and-classification procedure using publicly generated training data gives an IND-CPA attack against $\Pi'$ if $|\Pr[\mathrm{correct}]-1/2|$ is nonnegligible. Moreover, such a classifier yields adversaries against both the KEM and the PKE with the same advantage, so Algorithm~\ref{alg:xor_combiner_ind_cpa} can only succeed if neither component is IND-CPA.

\subsection{Considered Algorithms}\label{sec:KEMs}

To demonstrate the applicability of the deep learning-based IND-CPA testing approach to important and recent KEMs, we select the PKE component of the following algorithms from the submissions to the NIST PQC standardization process. 
\begin{inparaenum}
    \item \textbf{ML-KEM} (formerly called CRYSTALS-Kyber) \cite{Bos2018Kyber} is a lattice-based KEM and the primary PQC KEM selected for standardization by NIST \cite{kyber_standard}; and
    \item \textbf{HQC} \cite{hqc-aguilarmelchor:hal-01946880} (Hamming Quasi-Cyclic codes) is a code-based KEM, selected by NIST for standardization \cite{hqc_standard}.
\end{inparaenum}
For our purposes, we use the instances targeting NIST security level 1: ML-KEM-512 and HQC-128. We test the PKEs using a constant key, with randomness uniformly drawn per encryption. Each PKE encrypts messages of its native length: 256 bits for ML-KEM-512, and 128 bits for HQC-128.

 To demonstrate the classifiers' performance on detection of faulty algorithms, we also test on the following:
\begin{inparaenum}
    \setcounter{enumi}{2}
    \item The \textbf{RC4} stream cipher \cite{schneier2007applied}, in the setup used in the WEP protocol (24-bit IV drawn uniformly at random and 104-bit key; the message key is $\mathit{IV} \,\|\, K$) \cite{IEEEStd802.11-2024}. This is known to produce a biased keystream \cite{RC4_bias}. For the RC4 variant, we use 2-byte plaintexts, and the IVs are not given to the DNNs.
    \item \textbf{AES-CTR} and \textbf{AES-CBC} where the counter or IV is reset periodically.   In our case, we investigate the scenario where the reset happens either once every 1000 encryptions or once every 10,000 encryptions. For the AES variants we use plaintext of length 16 bytes.
    \end{inparaenum}

\subsection{Considered Hybrid KEM Setups}\label{sec:asym_alg}
For Algorithm~\ref{alg:xor_combiner_ind_cpa}, we reuse the PQC KEMs from Section~\ref{sec:KEMs}. To be able to apply Algorithm \ref{alg:xor_combiner_ind_cpa}, we require the second algorithm to be a cipher in the sense that we can set the plaintext. Since the scenario is key exchange, a PKE is selected; in this case, RSA. Using RSA allows for testing how the two PQC candidates perform combined with an IND-CPA algorithm, such as RSA using optimal asymmetric encryption padding (OAEP) \cite{rfc8017}, and how they combine with a non-IND-CPA algorithm, such as plain RSA without using any padding. 

We also test combining with plaintext, i.e., concatenating the shared secret or a random bit sequence of equal length to the KEM ciphertext. This is done to test a worst-case scenario, where the cipher is insecure.

Finally, we also test plain RSA combined with plaintext as well as RC4 combined with plaintext, with the purpose of testing two combinations of two insecure algorithms. 

 \subsection{Considered Cascade Encryption Setups (Symmetric Algorithms)}\label{sec:sym_alg}
For the cascade combiner, three modes of AES are selected, namely counter-mode (CTR), electronic codebook (ECB), and cipher block chaining (CBC) \cite{AES_modes}. Alongside the three AES variants, the stream cipher ChaCha20 \cite{bernstein2008chacha}, and DES in ECB mode are added \cite{DES}. This selection contains both algorithms that are IND-CPA (AES-CBC, AES-CTR, ChaCha20) and those that are not (AES-ECB, DES-ECB). 

We test every combination of outer and inner cipher possible, with a few exceptions: (i) No cipher is tested combined with itself; and (ii) The combination of AES-ECB and DES-ECB is not tested since it produces a deterministic cipher.

\subsection{Neural Network Architecture, Parameter Selection, and Evaluations}\label{sec:setup}

We select fully connected and residual architectures as two complementary hypothesis classes: fully connected networks impose no structural assumptions on the input, while residual networks, through skip connections, are biased toward detecting small deviations from uniformity~\cite{gohr2019}.

For all algorithms, KEM combinations, and cascade encryptions tested, four networks are trained, i.e., a small and a big fully connected network, and a small and a big residual network. The exact specifications are given in Table~\ref{tab:network_architectures}. All networks used each byte of the ciphertext as an input feature to the DNN.

\begin{table}[t]
    \centering
    \setlength{\tabcolsep}{3pt}
    \caption{Overview of network architectures tested.}
    \begin{tabular}{l|cccc}
         & \textbf{Small} & \textbf{Big}& \textbf{Small Res.} & \textbf{Big Res.} \\ \hline     
        Hidden layers & 2& 4& 3 & 5 \\
        Nodes per hidden layer& 100&600 & 100&600\\
        Skips &- &- & 1$\rightarrow$3 &1$\rightarrow$3, 3$\rightarrow$5\\    
        \end{tabular}

    \label{tab:network_architectures}
\end{table}

The networks (except for RC4 and RC4+plaintext) were trained on 500,000 samples per class, with 100,000 samples per class used as validation data to track network progress and facilitate early stopping. The final test of model performance was done on 100,000 samples per class. For the RC4 cases, the neural network was trained on 5,000,000 samples per class, validated on 500,000 samples per class, and finally tested on 5,000,000 samples per class. The training, testing, and validation data are generated with the same keying material, since the threat model is fixed-key. Each experiment, including data generation and neural network training, was repeated with new keys three times. 

Moreover, all networks used ReLU activation \cite{goodfellow2016deep} in the intermediate layers and a sigmoid activation function in the output layer. We train for at most 1000 epochs with early stopping, using a patience parameter of 100 and a minimum improvement of $ \delta_{ES} = 10^{-6}$. We only keep the models with the lowest validation loss. The training is done with a batch size of 1024, allowing for the DNNs to possibly learn weak patterns, if such patterns exist. We use an initial learning rate of $10^{-4}$ combined with plateau-based learning rate reduction (ReduceLROnPlateau), a common training heuristic where the learning rate is reduced by a factor of $0.5$ if the loss function does not improve by more than $\delta_{LR} = 10^{-5}$ over 20 consecutive epochs \cite{chollet2021deep}. We also bound the minimal learning rate to $10^{-7}$. 

Furthermore, we use a momentum of $0.9$ and use Nesterov Accelerated Gradient (NAG) \cite{goodfellow2016deep}, Glorot uniform weight initialization \cite{goodfellow2016deep}, and initialize the biases to zero. Training was performed on an NVIDIA GH200 96 GB GPU; convergence for a single network configuration typically required well under two hours. The code for data generation, training, and statistical evaluation is available at \texttt{https://github.com/ITSL-LiU/DNN-IND-CPA}.

\subsection{Statistical Testing}
To test if the DNNs perform better than random guessing, we use a two-sided binomial test. We use the significance level $\alpha = 0.01$, as it is a common choice in cryptographic applications; see, e.g., \cite[p.~1-4]{NIST800-22}. To account for the fact that we are performing multiple hypothesis tests, we use Bonferroni correction \cite{bland1995multiple}. This means rejecting the null hypothesis if $p<0.01/M$, where $M$ is the number of tests done.

As a baseline classifier, we compute the chi-squared statistic \cite{NIST800-22} comparing the byte counts within each ciphertext with those expected from uniformly distributed bytes. We predict the all-zero plaintext class if the statistic exceeds the upper $1\%$ critical value of $\chi^2_{255}$, and the uniformly random plaintext class otherwise. This is a heuristic classification rule with no trained parameters. It is therefore evaluated on independently generated data, with fresh keys for each of the three runs and the same number of samples as the DNN test sets.

\section{Main Results}\label{sec:results}
\subsection{Results of IND-CPA Testing for Single Algorithms}\label{sec:Kems_res}
We apply Algorithm \ref{alg:bce_cpa} with the network parameters described in Section \ref{sec:setup} to the algorithms listed in Section \ref{sec:KEMs}. The resulting classification accuracies for the DNN training are given in Table \ref{tab:results_single_alg}.

\begin{table*}[t]

\begin{center}
\caption{Accuracies and p-values for the application of Algorithm~\ref{alg:bce_cpa} to single (i.e., non-hybridized) algorithms. Each cell gives the mean accuracy $\pm$ the sample standard deviation (percentage points) over three independent runs and, in parentheses, the two-sided binomial p-value for the pooled runs.}\label{tab:results_single_alg}
{\small
\begin{tabular}{|l|c|c|c|c|c|}
\hline
{\bfseries Cryptographic} & {\bfseries Small} &{\bfseries Big} & {\bfseries Small} & {\bfseries Big}  &  \\
{\bfseries algorithm} &{\bfseries Net.} & {\bfseries Net.}  &{\bfseries Res.} & {\bfseries Res.} &$\chi^2$  \\ \hline
Plain RSA & 100.00$\pm$0.00\% & 100.00$\pm$0.00\% & 100.00$\pm$0.00\% & 100.00$\pm$0.00\% &  99.45$\pm$0.01\%\\
 & ($\dagger$) & ($\dagger$) & ($\dagger$) & ($\dagger$) & ($\dagger$) \\ \hline
RSA-OAEP & 49.96$\pm$0.11\% & 49.95$\pm$0.11\% & 49.98$\pm$0.07\% & 49.97$\pm$0.05\% & 50.00$\pm$0.01\%  \\
 & (0.52) & (0.41) & (0.74) & (0.64) & (0.99) \\ \hline
ML-KEM & 50.01$\pm$0.08\% & 49.92$\pm$0.10\% & 49.99$\pm$0.09\% & 50.08$\pm$0.13\% & 49.99$\pm$0.01\%  \\
 & (0.89) & (0.22) & (0.92) & (0.24) & (0.87) \\ \hline
HQC & 50.04$\pm$0.24\% & 49.97$\pm$0.09\% & 50.07$\pm$0.11\% & 49.96$\pm$0.10\% &  50.01$\pm$0.02\%  \\
 & (0.57) & (0.68) & (0.28) & (0.53) &  (0.88)\\ \hline
AES-CTR & 82.24$\pm$1.14\% & 99.85$\pm$0.02\% & 97.09$\pm$0.62\% & 99.88$\pm$0.01\% & 49.97$\pm$0.28\%   \\
Ctr. reset $10^3$ & ($\dagger$) & ($\dagger$) & ($\dagger$) & ($\dagger$) & (0.70) \\ \hline
AES-CTR & 53.80$\pm$0.51\% & 77.27$\pm$2.95\% & 54.68$\pm$0.49\% & 98.16$\pm$0.21\% & 50.03$\pm$0.08\%  \\
Ctr. reset $10^4$ & ($\dagger$) & ($\dagger$) & ($\dagger$) & ($\dagger$) & (0.68) \\ \hline
AES-CBC & 83.52$\pm$1.06\% & 99.85$\pm$0.01\% & 96.97$\pm$0.23\% & 99.87$\pm$0.01\% & 50.10$\pm$0.11\%  \\
IV reset $10^3$ & ($\dagger$) & ($\dagger$) & ($\dagger$) & ($\dagger$) & (0.14) \\ \hline
AES-CBC & 53.94$\pm$0.14\% & 74.98$\pm$1.10\% & 55.45$\pm$0.10\% & 98.10$\pm$0.19\% &  50.01$\pm$0.01\%\\
IV reset $10^4$ & ($\dagger$) & ($\dagger$) & ($\dagger$) & ($\dagger$) & (0.89) \\ \hline
RC4 & 50.12$\pm$0.02\% & 50.13$\pm$0.01\% & 50.12$\pm$0.02\% & 50.16$\pm$0.02\% & 50.00$\pm$0.00\%  \\
 & ($4.3\cdot 10^{-38}$) & ($3.3\cdot 10^{-45}$) & ($1.5\cdot 10^{-41}$) & ($5.3\cdot 10^{-69}$) & (0.86) \\ \hline
\end{tabular}
}
\end{center}
\vspace{2mm}
\noindent $\dagger$p-value below $10^{-100}$.

\end{table*}

First, we note that all DNNs achieve 100\% accuracy in classifying the plain RSA ciphertexts, demonstrating complete distinguishability of ciphertexts. The perfect accuracy is expected, as deterministic (plain) RSA is well-known to be vulnerable to IND-CPA attacks due to its deterministic encryption. Applying the DNN classifiers to RSA-OAEP, on the other hand, achieves performance that is statistically indistinguishable from random guessing, which is consistent with RSA-OAEP's proven IND-CPA security; see also the results in \cite{kim2025cryptanalysismachinelearningbased}. Plain RSA is, however, the only algorithm to produce an output that can be accurately classified by chi-squared tests, so the use of a DNN does not by itself provide additional insight here.

For ML-KEM and HQC, we see that all DNNs perform close to 50\% in accuracy, with p-values much larger than 0.01, failing to reject the null hypothesis that the networks perform no differently from random guessing. This is consistent with the algorithms being IND-CPA. 

For the symmetric algorithms with counter/IV resets, all four DNN setups learn a statistically significant classification advantage, even when the Bonferroni-corrected threshold for 45 tests, $\alpha =0.01/45$, is used, while the chi-squared test classifies the ciphertexts with an accuracy similar to random guessing. The flaw is visible across ciphertexts rather than within any single one. Chi-squared tests the byte distribution of an individual ciphertext, which remains uniform. Counter or IV reuse, however, restricts the all-zero class to a small recurring set of ciphertext values while random plaintexts remain spread across the full space, and the DNNs separate the classes by this difference in distributional support — a distinction that holds on held-out test data. For this case, there is a notable difference in accuracy between the different network architectures, where the larger networks and the networks containing layer skips have greater accuracy in ciphertext classification.

Finally, for RC4, all DNN classifiers learn to classify the ciphertexts with an advantage that is significantly different from random guessing, with p-values that are at most $4.3\cdot 10^{-38}$, well below the Bonferroni-corrected threshold of $\alpha = 0.01/45$, despite the small deviation from $50\%$ classification accuracy. This is consistent with the theoretical bias of approximately $2/256$ \cite{RC4_bias}, and the p-values provide strong evidence that the classifiers exploit genuine information leakage, rather than statistical fluctuations.

\subsection{Results of IND-CPA Testing for KEMs Combined with a PKE}\label{sec:comb_res}
\begin{table*}[t]

\begin{center}
\caption{Accuracies and p-values for the application of Algorithm~\ref{alg:xor_combiner_ind_cpa} to hybrids constructed using the XOR combiner. Each cell gives the mean accuracy $\pm$ the sample standard deviation (percentage points) over three independent runs and, in parentheses, the two-sided binomial p-value for the pooled runs.}\label{tab:results_combiner}
{\small
\begin{tabular}{|l|c|c|c|c|c|}

\hline
{\bfseries Algs.} & {\bfseries Small} &{\bfseries Big} & {\bfseries Small} & {\bfseries Big}  &  \\
{\bfseries } &{\bfseries Net.} & {\bfseries Net.}  &{\bfseries Res.} & {\bfseries Res.} &$\chi^2$  \\ \hline
ML-KEM + & 49.93$\pm$0.06\% & 50.08$\pm$0.13\% & 50.05$\pm$0.06\% & 50.08$\pm$0.16\% & 49.99$\pm$0.01\% \\
RSA-OAEP & (0.29) & (0.24) & (0.45) & (0.19) &  (0.87)\\ \hline
ML-KEM + & 50.03$\pm$0.05\% & 50.00$\pm$0.12\% & 50.00$\pm$0.11\% & 50.00$\pm$0.04\% & 50.01$\pm$0.01\% \\
Plain RSA & (0.59) & (0.97) & (0.97) & (0.99) & (0.87) \\ \hline
ML-KEM + & 49.99$\pm$0.17\% & 49.88$\pm$0.13\% & 50.05$\pm$0.10\% & 49.93$\pm$0.10\% &50.01$\pm$0.01\%\\
Plaintext & (0.93) & (0.061) & (0.41) & (0.26) & (0.86) \\ \hline

HQC + & 50.00$\pm$0.20\% & 50.06$\pm$0.08\% & 49.94$\pm$0.03\% & 49.88$\pm$0.08\% & 49.98$\pm$0.03\%\\
RSA-OAEP & (0.99) & (0.35) & (0.37) & (0.065) & (0.82) \\ \hline
HQC + & 49.93$\pm$0.07\% & 50.02$\pm$0.18\% & 49.96$\pm$0.15\% & 50.02$\pm$0.05\% & 49.99$\pm$0.03\%\\
Plain RSA & (0.27) & (0.74) & (0.53) & (0.74) & (0.89) \\ \hline
HQC + & 50.10$\pm$0.05\% & 49.90$\pm$0.10\% & 50.02$\pm$0.14\% & 49.99$\pm$0.17\% & 50.00$\pm$0.01\% \\
Plaintext & (0.14) & (0.11) & (0.72) & (0.88) & (1.0) \\ \hline

Plain RSA + & 49.91$\pm$0.17\% & 50.05$\pm$0.08\% & 49.96$\pm$0.19\% & 50.07$\pm$0.13\% & 49.99$\pm$0.02\% \\
Plaintext & (0.16) & (0.43) & (0.54) & (0.29) & (0.83) \\ \hline

RC4 + & 50.04$\pm$0.01\% & 50.06$\pm$0.04\% & 50.05$\pm$0.01\% & 50.06$\pm$0.00\% & 50.19$\pm$0.00\% \\
Plaintext & ($1.4\cdot 10^{-5}$) & ($2.0\cdot 10^{-11}$) & ($3.6\cdot 10^{-8}$) & ($5.3\cdot 10^{-10}$) &($1.7\cdot 10^{-98}$) \\ \hline
\end{tabular}
}
\end{center}

\end{table*}

We apply Algorithm~\ref{alg:xor_combiner_ind_cpa} with the DNN parameters described in Section~\ref{sec:setup} to the combinations described in Section~\ref{sec:asym_alg}, the results can be found in table \ref{tab:results_combiner}. For the combinations of PQC KEMs with PKEs, no result is statistically significant, neither at the significance level $\alpha=0.01$ nor at the Bonferroni-corrected threshold of $\alpha=0.01/40$.

These results indicate the robustness of the XOR combiner, as the DNN classifier does not learn any patterns that allow it to classify the ciphertexts better than random guessing. Notably, the accuracies remain close to 50\% even when an insecure component algorithm, such as plain RSA or plaintext, is used. For constructions covered by the combiner theory from~\cite{Giacon2018Combiners}, as discussed in Section~\ref{subsec:HybridEncryption}, this is consistent with the theoretical guarantee that security is inherited from the stronger component.

One combination of two weak algorithms, namely RC4 + plaintext, resulted in ciphertext pairs that the DNNs could classify more accurately than random guessing, by exploiting the biases in the RC4 keystream alongside the known plaintext. This weakness was also detected using chi-squared tests, but it demonstrates that the DNN adversary is able to exploit cross-component weaknesses in certain cases.

Despite both plain RSA and plaintext being insecure, the four networks classify this pair with average accuracies in the range 49.91\%-50.07\% (p-values 0.16-0.54), which is indistinguishable from random guessing. An adversary tasked with classifying the hybrid ciphertext $c_{\mathrm{RSA}}\|m$, who also has access to the RSA public key, could encrypt $m$ and compare the result to $c_{\mathrm{RSA}}$, achieving accuracy $1-2^{-\ell-1}$ for equally likely classes and an independent uniform $\ell$-bit comparison value. Our DNN adversary operates under a more restrictive model, as it lacks such oracle access. This result thus does not imply security for this pair of algorithms but illustrates that ciphertext distinguishability can depend on factors such as knowledge of public parameters or oracle access, and highlights a limitation of the DNN approach to IND-CPA analysis. 

\subsection{Results of IND-CPA Testing of Cascade Encryption}\label{sec:cascade_res}

\begin{table*}[t]

\begin{center}
\caption{Accuracies and p-values for the application of Algorithm~\ref{alg:bce_cpa} to cascade encryption. Each cell gives the mean accuracy $\pm$ the sample standard deviation (percentage points) over three independent runs and, in parentheses, the two-sided binomial p-value for the pooled runs. Algorithms are noted as inner cipher + outer cipher.}\label{tab:results_cascades}
{\small
\begin{tabular}{|l|c|c|c|c|c|}
\hline
{\bfseries Algs.} & {\bfseries Small} &{\bfseries Big} & {\bfseries Small} & {\bfseries Big}  &   \\
{\bfseries } &{\bfseries Net.} & {\bfseries Net.}  &{\bfseries Res.} & {\bfseries Res.} &$\chi^2$ \\ \hline
AES-CBC & 49.90$\pm$0.12\% & 50.13$\pm$0.05\% & 50.07$\pm$0.04\% & 50.18$\pm$0.10\% & 50.01$\pm$0.02\% \\
AES-CTR & (0.14) & (0.037) & (0.31) & (0.0045) & (0.87)  \\ \hline
AES-CBC & 50.01$\pm$0.08\% & 50.05$\pm$0.11\% & 50.03$\pm$0.04\% & 50.07$\pm$0.12\% &49.99$\pm$0.01\%  \\
AES-ECB & (0.88) & (0.48) & (0.63) & (0.26) &(0.85) \\  \hline
AES-CBC & 49.92$\pm$0.09\% & 49.99$\pm$0.10\% & 49.94$\pm$0.03\% & 50.03$\pm$0.08\% & 50.00$\pm$0.04\%  \\
ChaCha20 & (0.20) & (0.92) & (0.36) & (0.67) & (0.96) \\ \hline
AES-CBC & 50.06$\pm$0.04\% & 50.15$\pm$0.07\% & 49.91$\pm$0.02\% & 50.12$\pm$0.09\% & 50.07$\pm$0.02\% \\
DES & (0.39) & (0.017) & (0.17) & (0.073) &(0.27) \\ \hline

AES-CTR & 50.08$\pm$0.06\% & 49.98$\pm$0.01\% & 50.02$\pm$0.08\% & 49.87$\pm$0.12\% & 50.05$\pm$0.01\% \\
AES-CBC & (0.19) & (0.76) & (0.72) & (0.046) & (0.47) \\ \hline
AES-CTR & 50.05$\pm$0.16\% & 50.05$\pm$0.04\% & 50.06$\pm$0.16\% & 50.00$\pm$0.07\% & 50.00$\pm$0.02\% \\
AES-ECB & (0.45) & (0.42) & (0.38) & (0.96) &  (0.99)\\ \hline
AES-CTR & 50.06$\pm$0.13\% & 50.02$\pm$0.08\% & 49.93$\pm$0.13\% & 49.99$\pm$0.10\% & 50.01$\pm$0.01\% \\
ChaCha20 & (0.35) & (0.75) & (0.25) & (0.90) & (0.92) \\ \hline
AES-CTR & 49.95$\pm$0.06\% & 49.98$\pm$0.03\% & 50.00$\pm$0.03\% & 50.03$\pm$0.11\% & 50.00$\pm$0.02\% \\
DES & (0.44) & (0.73) & (0.99) & (0.67) & (0.97)  \\ \hline

AES-ECB & 49.97$\pm$0.15\% & 50.00$\pm$0.08\% & 50.03$\pm$0.11\% & 49.97$\pm$0.03\% & 50.00$\pm$0.02\% \\
AES-CBC & (0.62) & (0.94) & (0.60) & (0.60) & (0.99) \\ \hline
AES-ECB & 49.92$\pm$0.07\% & 50.01$\pm$0.16\% & 49.89$\pm$0.10\% & 50.01$\pm$0.17\% &50.00$\pm$0.03\%   \\
AES-CTR & (0.23) & (0.92) & (0.078) & (0.93) & (0.95)  \\ \hline
AES-ECB & 49.96$\pm$0.08\% & 50.15$\pm$0.05\% & 50.08$\pm$0.19\% & 49.94$\pm$0.03\% &  49.96$\pm$0.02\%  \\
ChaCha20 & (0.58) & (0.024) & (0.22) & (0.38) & (0.50) \\ \hline

ChaCha20 & 49.92$\pm$0.08\% & 49.95$\pm$0.11\% & 50.16$\pm$0.04\% & 49.99$\pm$0.03\% & 50.01$\pm$0.02\%  \\
AES-CBC & (0.20) & (0.43) & (0.013) & (0.88) & (0.92) \\ \hline
ChaCha20 & 49.92$\pm$0.08\% & 49.92$\pm$0.12\% & 50.04$\pm$0.06\% & 49.93$\pm$0.13\% &50.00$\pm$0.01\%  \\
AES-CTR & (0.21) & (0.24) & (0.50) & (0.31) & (1.0) \\ \hline
ChaCha20 & 49.97$\pm$0.09\% & 50.06$\pm$0.13\% & 49.97$\pm$0.11\% & 50.05$\pm$0.02\% &  49.99$\pm$0.01\% \\
AES-ECB & (0.67) & (0.35) & (0.66) & (0.45) & 0.88 \\ \hline
ChaCha20 & 49.99$\pm$0.03\% & 50.05$\pm$0.08\% & 50.06$\pm$0.05\% & 49.99$\pm$0.17\% & 50.01$\pm$0.04\%  \\
DES & (0.82) & (0.41) & (0.33) & (0.82) & (0.90) \\ \hline

DES & 50.04$\pm$0.08\% & 50.02$\pm$0.14\% & 49.98$\pm$0.13\% & 50.01$\pm$0.13\% & 50.00$\pm$0.04\%  \\
AES-CBC & (0.54) & (0.76) & (0.80) & (0.85) & (0.96) \\ \hline
DES & 49.88$\pm$0.17\% & 49.96$\pm$0.06\% & 50.00$\pm$0.02\% & 49.94$\pm$0.11\% & 49.99$\pm$0.03\% \\
AES-CTR & (0.057) & (0.54) & (0.99) & (0.36) & (0.84) \\ \hline
DES & 49.96$\pm$0.08\% & 50.02$\pm$0.22\% & 50.01$\pm$0.11\% & 49.97$\pm$0.13\% &  49.99$\pm$0.04\% \\
ChaCha20 & (0.53) & (0.79) & (0.92) & (0.62) & (0.94) \\ \hline
\end{tabular}
}
\end{center}
\end{table*}

The results of applying Algorithm \ref{alg:bce_cpa} to the cascaded symmetric algorithms can be found in Table \ref{tab:results_cascades}. As can be seen in the table, no neural network achieved a significant classification advantage over the Bonferroni-corrected threshold of $\alpha=0.01/90$. These results are consistent with the ciphertext indistinguishability of cascade symmetric ciphers, as the neural classifier cannot classify the ciphertexts better than random guessing. Notably, the combinations remain resistant to the DNN classifier when the cascade is done with one deterministic non-IND-CPA cipher, such as AES-ECB or DES. In the cascade experiment, our results indicate no difference between the network architectures.
\section{Conclusion}\label{sec:conclusion}
We applied the DNN IND-CPA classification framework to empirically test the IND-CPA property of both the underlying PKE components used to construct two PQC algorithms and symmetric algorithms subject to IV or counter reset. For the PQC PKEs, none of our classifiers achieves an accuracy that is significantly different from random guessing, which is consistent with the algorithms being IND-CPA. For the symmetric algorithms under IV or counter reset, the classifiers showed a statistically significant advantage that was not detected by the chi-squared test.

Given sufficient training data (5,000,000 samples per class), the classifiers also detected the keystream bias of RC4 in the WEP setting, with accuracies of 50.12--50.16\% and p-values below $10^{-37}$, showing that the framework can identify subtle weaknesses.

We also presented an extension of the framework to evaluate ciphertext distinguishability for hybrid constructions using the XOR combiner under the considered DNN adversary model. We also demonstrated that the methodology of using a DNN to model IND-CPA games can also be applied to cascade encryption. Our experiments provided no evidence that cascade encryption degrades IND-CPA security, since in all tested configurations, the classifiers achieved accuracies indistinguishable from random guessing whenever at least one component algorithm is IND-CPA. 

Thus, our results demonstrate the versatility in using DNN classifiers as empirical tools for testing the IND-CPA security of both individual algorithms and composite constructions, including hybrid KEMs and cascade encryption. Moreover, we discussed that the DNN adversary is strictly weaker than a standard IND-CPA adversary, so a negative result does not constitute a proof of security. For future work, it is therefore of high interest to develop DNN classifiers that can incorporate factors such as public key or encryption oracle access.


\section*{Acknowledgment}
This research has been partially supported by the Swedish Foundation for Strategic Research (SSF) under grant ID24-0087 and the German Federal Ministry of Research, Technology and Space (BMFTR) 6GEM+ Transfer Hub under the Grants 16KIS2412 and 16KISS005. The training of the DNNs was enabled by resources provided by the National Academic Infrastructure for Supercomputing in Sweden (NAISS).
\bibliographystyle{IEEEtran}
\bibliography{references}
\end{document}